\documentclass[a4paper]{jpconf}
\usepackage{graphicx}
\usepackage{subfigure}
\usepackage{mathrsfs}
\usepackage{epsfig}
\newcommand{\ot}{\otimes}
\newcommand{\ket}[1]{|{#1}\rangle}
\newcommand{\bra}[1]{\langle{#1}|}
\newcommand{\bkt}[2]{\langle{#1}|{#2}\rangle}
\begin{document}
\title{Enhancing the
sensitivity of amplification of rotation velocity in Sagnac's interferometer with biased weak measurement }

\author{Jing-Hui Huang$^{1}$, Xiang-Yun Hu$^{1*}$, Xue-Ying Duan$^{2,3,4}$ and Guang-Jun Wang$^{2,3,4}$}
\address{1. Institute of Geophysics and Geomatics, China University of Geosciences, Wuhan 430074, China}
\address{2. School of Automation, China University of Geosciences, Wuhan 430074, China}
\address{3. Hubei Key Laboratory of Advanced Control and Intelligent Automation for Complex Systems, Wuhan 430074, China}
\address{4. Engineering Research Center of Intelligent Technology for Geo-Exploration, Ministry of Education}
\ead{jinghuihuang@cug.edu.cn,xyhu@cug.edu.cn}

\begin{abstract}
Recently, biased weak measurement(BWM) has shown higher precision than both conventional measurement and standard weak measurement(SWM) in optical metrology. In this paper, we propose a scheme of detecting rotation velocity in Sagnac's interferometer with BWM. In particular, BWM employs an additional reduction of photons in the post-selection by introducing a pre-coupling, and the remaining photons have been shown to be extremely sensitive to the estimated parameter. In addition, our numerical results show that the scheme with BWM can obtain a higher sensitivity than the scheme with SWM.
\end{abstract}

\section{Introduction}
Standard weak measurement(SWM) proposed by Aharonov, Albert and Vaidman in 1988\cite{AAV} plays an important role in fields of modern science and industry, where information is gained by weakly coupling the probe to the quantum system. Normally, SWM is characterized by preparation for the measuring system, a weak interaction between the measuring device(probe) and measured system, post-selection on the system, and a projective measurement on the measuring device to read out the measurement result. Numerous studies show that the quantum measurements with weak value amplification(WVA) can beat the classical schemes with high sensitivity and good robustness, such as
amplification of angle rotations\cite{PhysRevLett.112.200401}, longitudinal velocity shifts\cite{2013Weak}, frequency shifts\cite{PhysRevA.82.063822}, the Goos-Hanchen shift\cite{Santana:19}, rotation velocity\cite{Huang2021}, linear velocity\cite{Huang2021-2} and the photonic spin Hall effect\cite{2021Weak}.

Recently, biased weak measurement(BWM) has shown higher precision than both conventional measurement and standard weak measurement(SWM) in optical metrology\cite{BWM2016,2021Yin}. BWM employs an additional reduction of photons in the post-selection by introducing a pre-coupling, and the remaining photons have been shown to be extremely sensitive to the estimated parameter. In this work, we propose a scheme to enhance the sensitivity of amplification of rotation velocity in Sagnac's interferometer with BWM.

Our paper is organized as follows. In section 2, we briefly introduce the scheme of detecting rotation velocity in Sagnac’s interferometer with SWM. In section 3, a modified scheme with BWM is introduced to enhance the sensitivity of amplification of rotation velocity in Sagnac's interferometer. In section 4, we give the numerical simulation of these schemes and analyze the results. We conclude in section 5  and possible extensions for our work.

\section{Framework of SMW for detecting rotation velocity in Sagnac’s interferometer}

The basic principle of Sagnac's interferometer is given in Ref\cite{Post1967}. A beam of light from the same light source is split into two beams, which travel in the same loop in opposite directions for a circle and then meet. When the whole interferometer is set in rotation with an angular rate of $\Omega$ rad/sec, a fringe shift $\Delta z$ with respect to the fringe position for the stationary is observed, which is given by the form
\begin{eqnarray}
\label{deltz}
\Delta z = \frac{4\Omega S}{\lambda_{0} c}
\end{eqnarray}
where S is the area enclosed by the light path. The vacuum wavelength is $\lambda_{0}$ and the free space velocity of light is c. Normally, in order to detect the small rotation signal, the larger S is needed. For example, the $4 \times 4 m$ ring laser system\cite{Heiner2010} that was installed at the geodetic observatory Wettzell, SE-Germany to detect the earthquake-induced ground motions far from seismic sources.

\begin{figure}[t]
\centering
\includegraphics[width=0.79\textwidth]{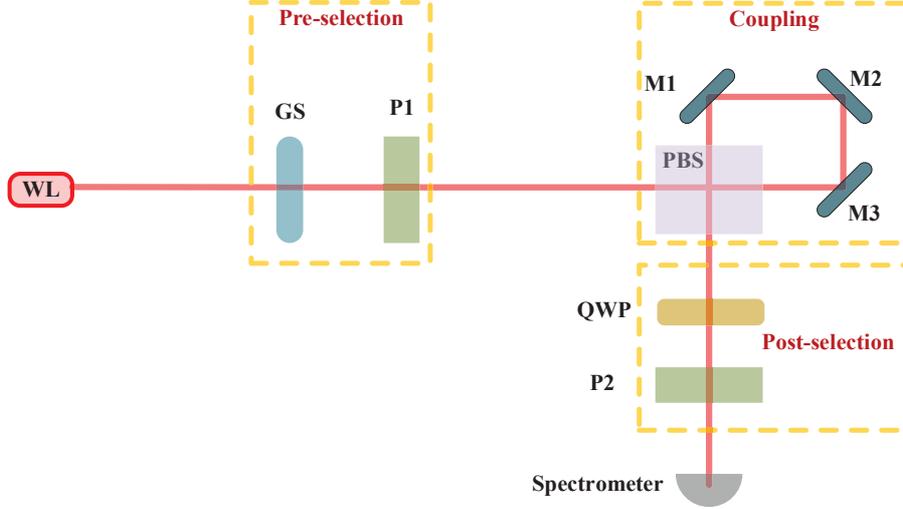}
\vspace*{-5mm}
\caption{The scheme of SMW. A white laser(WL) as the light source; GF: Gaussian filter with the central wavelength $\lambda_{0}$ and the full width at half maximum $\Delta  \lambda^{2}$; P1 and P2: Polarizers with optical axes set $45^{o}$ and $\varphi-45^{o}$, respectively; PBS: Polarized beam splitter where the beam is injected into a Sagnac's interferometer with horizontally polarized component $\ket{H}$ and vertically polarized component $\ket{V}$; M1, M2 and M3: common mirrors; QWP: Quarter wave plate with optical axis set at  $-45^{o}$. }
\label{scheme1}
\end{figure}

SMW is an innovative method to determine small physical quantities or optical phases. The scheme of SMW for detecting the Sagnac phase $ \Delta \Phi_{m}=2\pi \Delta z$ caused by the rotation velocity in Sagnac's interferometer is shown in Figure \ref{scheme1}. SMW is characterized by state preparation, a weak perturbation, and postselection. We prepare the initial state $\ket{\Phi_{i}}$ of the system and $\ket{\Psi_{i}}$ of the probe. After a certain interaction between the system and the probe, we post-select a system state $\ket{\Phi_{f}}$ and obtain information about a physical quantity $\hat{A}$ from the probe wave function by the weak value
\begin{eqnarray}
\label{weak_value}
A_{w}:=\frac{\bra{\Phi_{f}}\hat{A}\ket{\Phi_{i}}}{\bkt{\Phi_{f}}{\Phi_{i}}},
\end{eqnarray}
which can generally be a complex number. More precisely, the shifts of the position and momentum in the probe wave function are given by the real and imaginary parts of the weak value $A_{w}$, respectively. We can easily see from Eq. \Ref{weak_value} that when $\ket{\Phi_{i}}$ and $\ket{\Phi_{f}}$ are almost orthogonal, the absolute value of the weak value can be arbitrarily large. In our scheme, the Sagnac effect introduces an additional optical phase $ \Delta \Phi_{m}$ between two orthogonal polarization components $\ket{H}$ and $\ket{V}$. In particular, it is efficient to detect the small optical delay $\tau=\frac{\Phi_{m}}{cp_{0}}$ with the appropriate pre-selected and post-selected state, where the photons momentum is $p_{0}=\frac{2 \pi}{\lambda_{0}}$, and $\lambda_{0}$ is corresponding the central wavelength of incident light. Theoretically, the coupling strength $g=c \tau=\frac{\Phi_{m}}{p_{0}}$ can be estimated by the interaction between the system and the probe, which is initialized to $\int dp \ket{\Psi_{i}(p)}$ and is assumed to have a Gaussian profile with mean value $p_{0}$ and variance $(\delta p)^{2}$\cite{2021Yin}. 

The interaction Hamiltonian is:
\begin{eqnarray}
\label{Hamiltonian}
H=g\delta(t-t_{0})\hat{A} \ot \hat{p},
\end{eqnarray}
where $\hat{A}=\ket{H} \bra{H}-\ket{V} \bra{V} $ is the system operator, and $\hat{p}$ is the probe momentum operator conjugate to the position operator $\hat{q}$. We have taken the interaction to be impulsive at time $t=t_{0}$ for simplicity. The time evolution operator becomes $ e^{-ig\hat{A} \ot \hat{p}}$. 

In the Scheme of SMW, the pre-selection $\ket{\Phi_{i}}$ is obtained by the P1:
\begin{eqnarray}
\label{inter_sy_initial}
\ket{\Phi_{i}}= \frac{1}{\sqrt{2}} (\ket{H}+ \ket{V}) \, . 
\end{eqnarray}
For the post-selection in our scheme, we chose the combination of a quarter-wave plate and a polarizer, which has been successfully applied as the post-selection in Ref.\cite{Qiu2017,Zhu2021}. The post-selection $\ket{\Phi_{f}}$ of the system takes the following form:
\begin{equation}
\ket{\Phi_{f}} \simeq \frac{1}{\sqrt{2}}\left(\begin{array}{cc}
1 & -i \\
-i & 1
\end{array}\right)\left(\begin{array}{c}
\cos (\varphi-\pi / 4) \\
\sin (\varphi-\pi / 4)
\end{array}\right) =\frac{1}{\sqrt{2}}[\exp (-i \varphi)|H\rangle-\exp (i \varphi)|V\rangle] \,,
\end{equation}
where $\varphi$ is the angular between the optical axes of QWP and P2. After postselection, the probe state becomes 
\begin{eqnarray}
\label{inter_prob_final}
\ket{\Psi_{f}} &=&\bra{\Phi_{f}}e^{-ig\hat{A}\ot \hat{p}}\ket{\Phi_{i}}\ket{\Psi_{i}} \nonumber \\
&=&\bra{\Phi_{f}}\left[ 1-ig\hat{A}\ot \hat{p}\right]\ket{\Phi_{i}}\ket{\Psi_{i}}+O(g^{2}) \nonumber \\
&=&\bkt{\Phi_{f}}{\Phi_{i}}\left[ 1-igA_{w}\hat{p}\right]\ket{\Psi_{i}}+O(g^{2}) \nonumber \\
&=&\bkt{\Phi_{f}}{\Phi_{i}}e^{-igA_{w}\hat{p}}\ket{\Psi_{i}}+O(g^{2}).
\end{eqnarray}
Combing weak value (\ref{weak_value}) and equation (\ref{inter_prob_final}), the spectrum shift of SWM is amplified by the weak value $A_{w}=\frac{\bra{\Phi_{f}}\hat{A}\ket{\Phi_{i}}}{\bkt{\Phi_{f}}{\Phi_{i}}}=i {\rm cot}\varphi$. And the distribution of $p$ in the scheme of SMW is given as:
\begin{eqnarray}
\label{final_probe_SWM}
|{\bkt{\Psi_{f}}{\Psi_{f}}}|^{2}_{SWM}=
{\rm sin^{2}} (gp+\varphi) |{\bkt{\Psi_{i}}{\Psi_{i}}}|^{2}.
\end{eqnarray}
Note that the coupling strength $g$ can be estimated through the shift of the mean value of $p$, which can be calculated as follow:
\begin{eqnarray}
\label{shift_SMW}
\delta p_{SWM}=2g (\delta p)^{2}  {\rm cot}\varphi\simeq\frac{16\pi   (\delta p)^{2}}{\lambda_{0} c S \varphi}  \Omega
\end{eqnarray}
where the minute rotation velocity $\Omega$ can be amplified by choosing orthogonal quantum states of $\ket{\Phi_{i}}$ and 
$\ket{\Phi_{f}}$. Theoretically, the smaller $\varphi$ is, the shifts $\delta p_{SWM}$ is bigger, and the price of the amplification is the lower possibility for post-selecting the photons.

\section{Framework of BMW for enhancing sensitivity of amplification of rotation velocity}

\begin{figure}[t]
\centering
\includegraphics[width=0.79\textwidth]{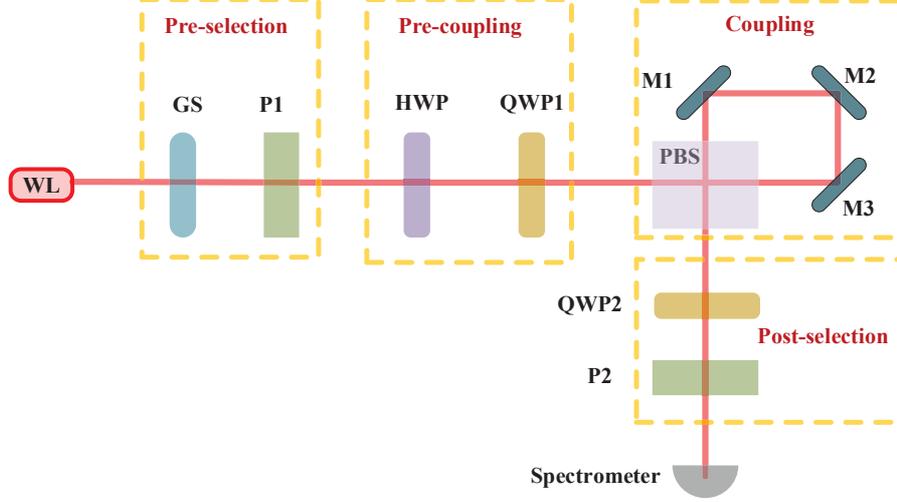}
\vspace*{-5mm}
\caption{The scheme of BMW. A white laser(WL) as the light source; GF: Gaussian filter with the central wavelength $\lambda_{0}$ and the full width at half maximum $\Delta  \lambda^{2}$; P1 and P2: Polarizers with optical axes set $45^{o}$ and $\varphi-45^{o}$, respectively;  HWP(Half wave plate) and QWP1(Quarter wave plate) which introduce a biased phase; PBS: Polarized beam splitter where the beam is injected into a Sagnac's interferometer with horizontally polarized component $\ket{H}$ and vertically polarized component $\ket{V}$; M1, M2 and M3: common mirrors; QWP2: Quarter wave plate with optical axis set at  $-45^{o}$. }
\label{scheme2}
\end{figure}

The BMW was proposed by Zhang etc.\cite{BWM2016}, where an extra bias phase can significantly improve the sensitivity of measuring small longitudinal phase change. This bias phase can be introduced by a pre-coupling process. Together with a special postselection, destructive interference can be observed in both the time and frequency domain. Thus, the scheme of BMW for enhancing the sensitivity of amplification of rotation velocity in Sagnac’s interferometer with SMW is shown in Figure \ref{scheme2}. Note that the main difference from the scheme of SMW(Figure \ref{scheme1}.) is an additional step to bias the probe before the coupling process. 

In the scheme of BMW,  a pre-determined time delay $\psi_{pre}/c$ is introduced by HWP and QWP1 between the horizontally polarized component $\ket{H}$ and vertically polarized component $\ket{V}$, where the $\psi_{pre}$ ought to satisfy the following relationships:
\begin{eqnarray}
\label{baised_condition}
p_{0}\psi_{pre}+\varphi=m \pi\, ,  \,\,{\rm where }\,\, m\,\, {\rm is }
\,\, {\rm an }\,\, {\rm integer }
\end{eqnarray}

Thus, the corresponding distribution of the post-selected probe state is given by:
\begin{eqnarray}
\label{final_probe_BWM}
|{\bkt{\Psi_{f}}{\Psi_{f}}}|^{2}_{BWM}=
{\rm sin^{2}} (p(g+\psi_{pre})+\varphi) |{\bkt{\Psi_{i}}{\Psi_{i}}}|^{2}=
{\rm sin^{2}} (pg) |{\bkt{\Psi_{i}}{\Psi_{i}}}|^{2}\,\,, m=0.
\end{eqnarray}

Finally, the mean value shift in the scheme of BWM is calculated as follows:
\begin{eqnarray}
\label{shift_BMW}
\delta p_{BWM}=2g ( p_{0})^{2}  {\rm cot}\varphi\simeq\frac{16\pi   (p_{0})^{2}}{\lambda_{0} c S \varphi}  \Omega.
\end{eqnarray}

\section{Numerical simulation and analysis of results}

\begin{figure*}[htp!]
	\centering
\subfigure
{
	\vspace{-0.2cm}
	\begin{minipage}{8.5cm}
	\centering
	\centerline{\includegraphics[scale=0.62,angle=0]{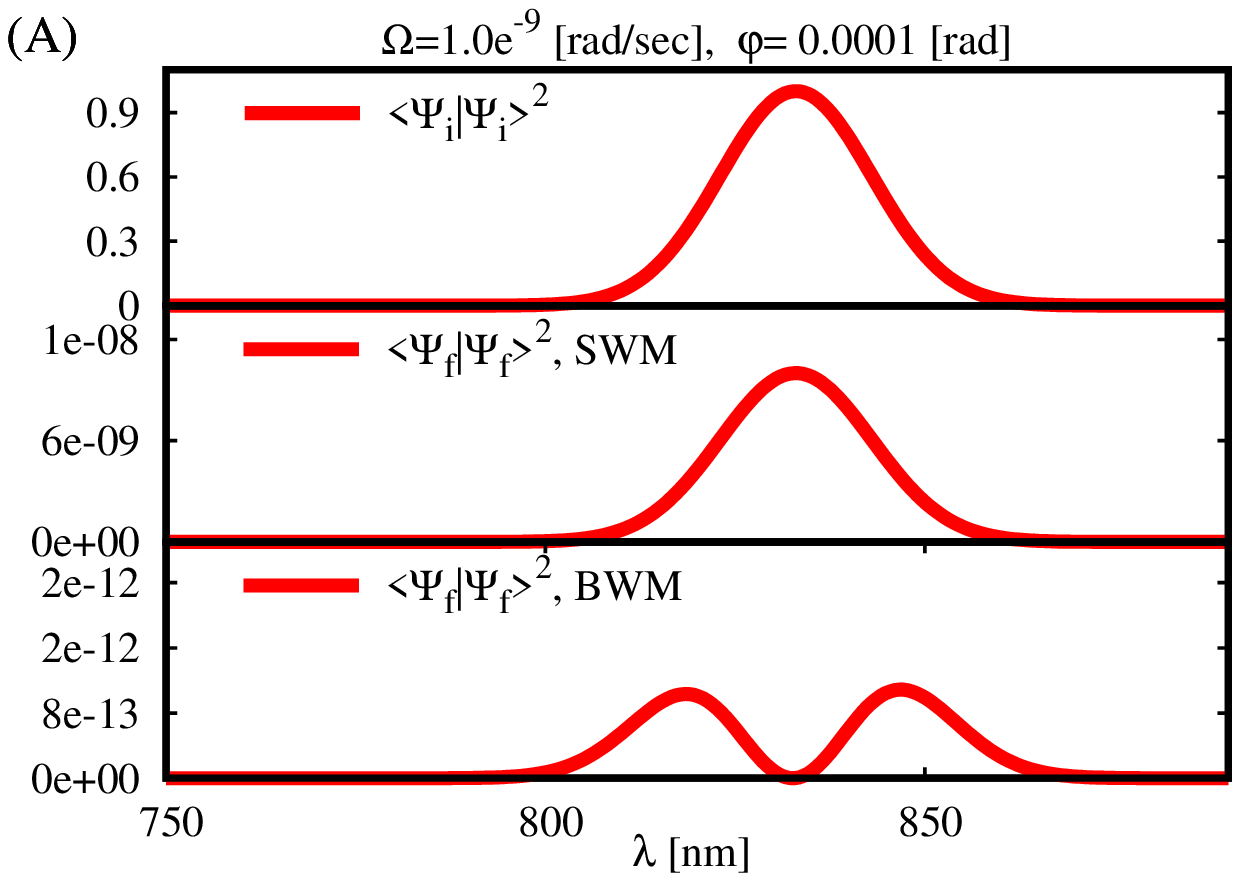}}
	\end{minipage}
}
\vspace{-0.2cm}
\subfigure
{
	\begin{minipage}{6.4cm}
	\centering
	\centerline{\includegraphics[scale=0.62,angle=0]{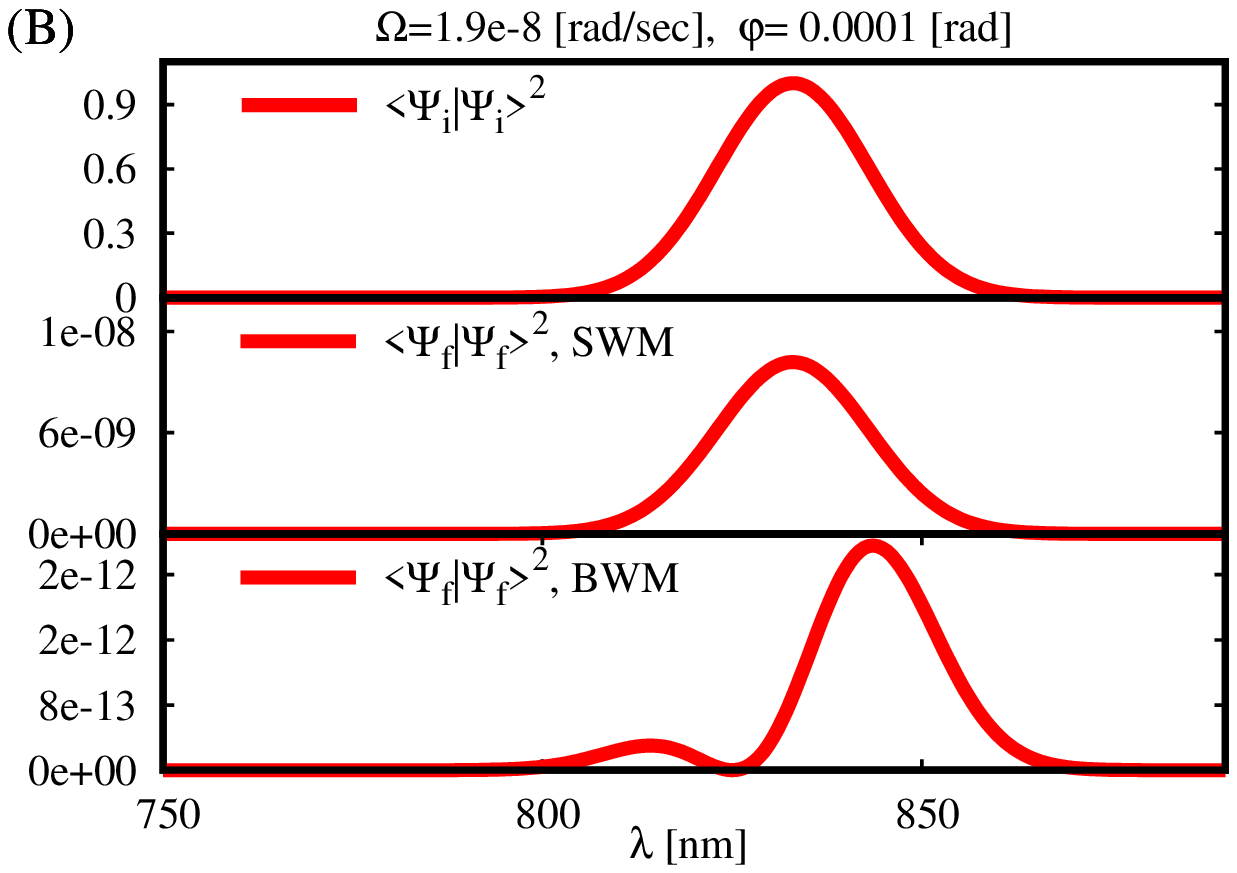}}
	\end{minipage}
}
\vspace{-0.2cm}

\subfigure
{
	\begin{minipage}{8.5cm}
	\centering
	\centerline{\includegraphics[scale=0.62,angle=0]{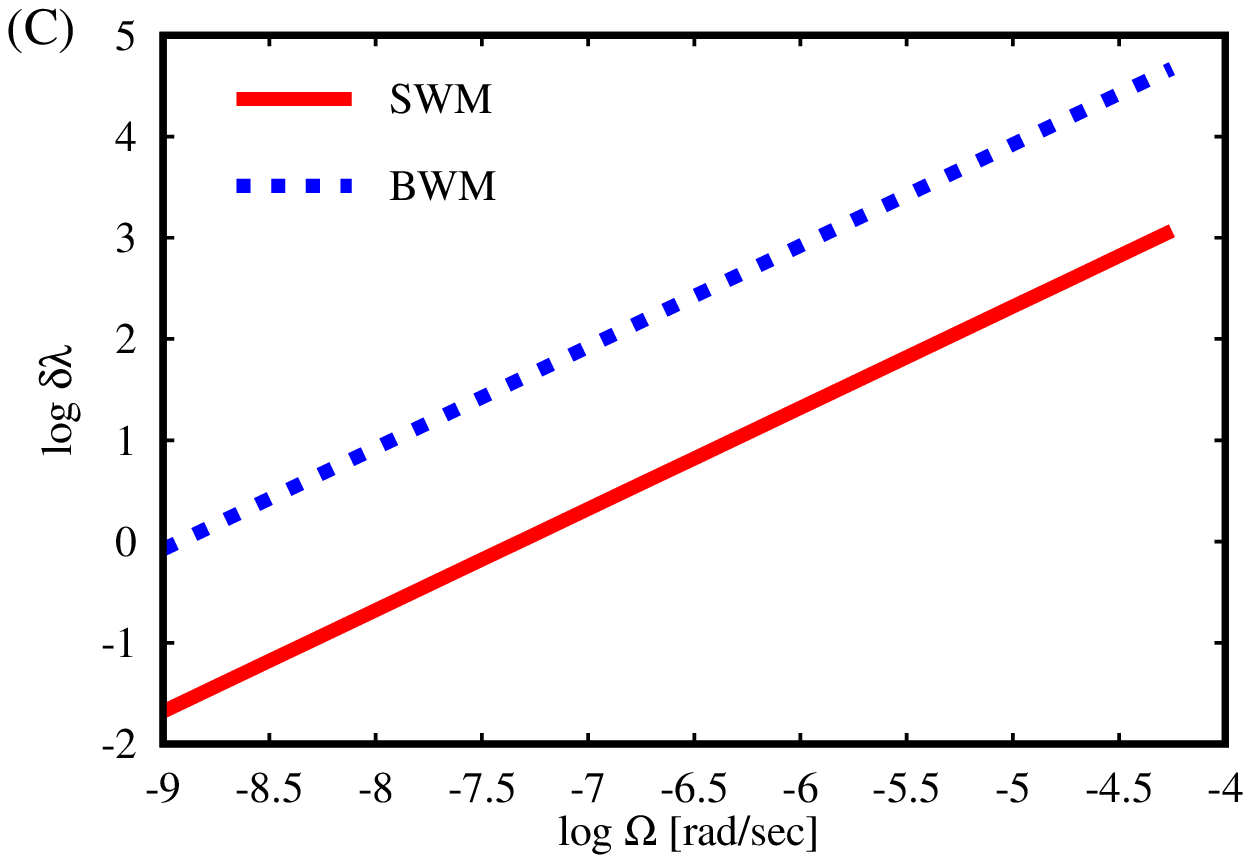}}
	\end{minipage}
}
\subfigure
{
	\begin{minipage}{6.4cm}
	\centering
	\centerline{\includegraphics[scale=0.62,angle=0]{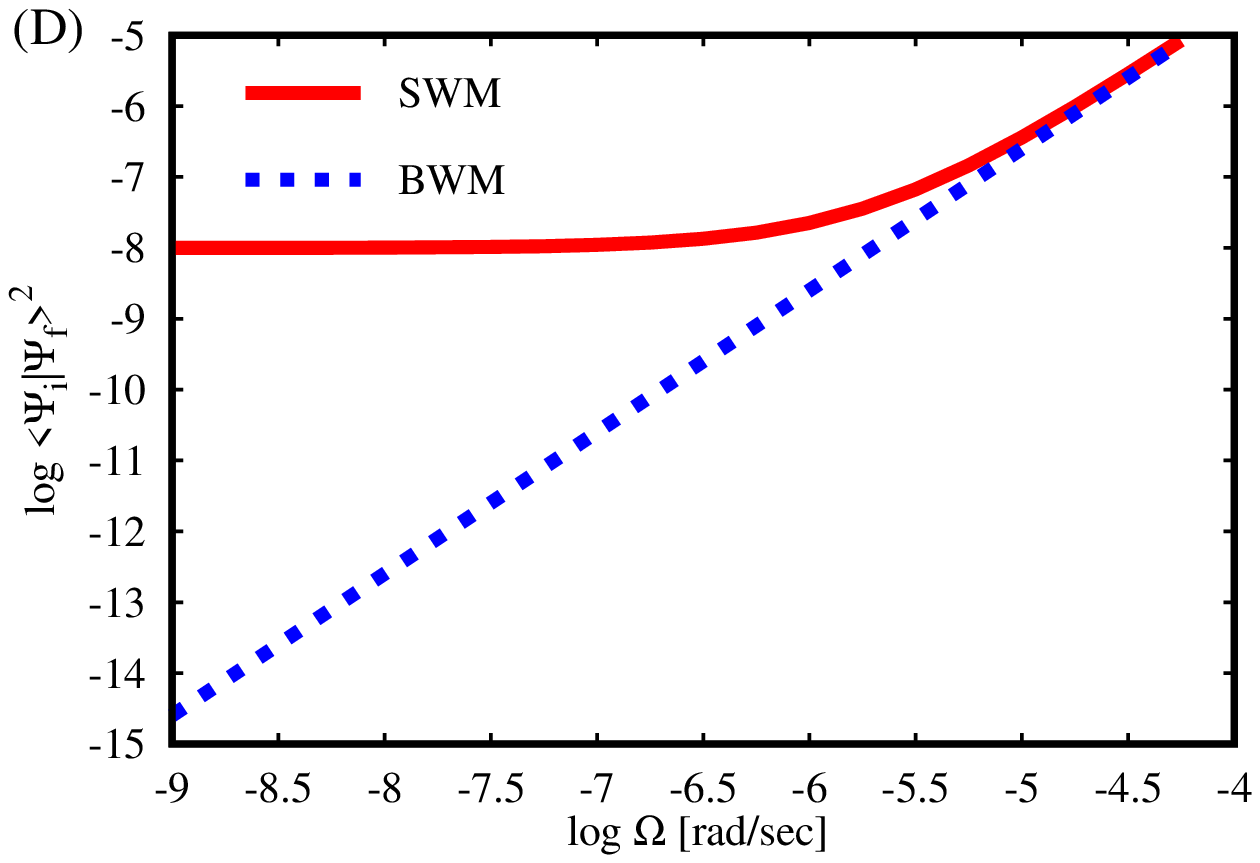}}
	\end{minipage}
}
\vspace{-0.2cm}

\vspace*{0mm} \caption{\label{result_numberical}The numerical results of the simulations of SWM and BWM.  }
\end{figure*}

In both the scheme of SWM and the scheme of BWM, the minute rotation velocity $\Omega$ are determined from the corresponding spectrum shifts. Considering the relationship of the photon momentum $p$ and the photon wavelength $\lambda$: $p=2\pi / \lambda$, $\delta p=-2\pi \delta \lambda/ \lambda^{2}$ and $\Delta p=-2\pi \Delta \lambda/ \lambda^{2}$, we obtain the shifts of the photon wavelength $\Delta \lambda_{SWM}$ and $\Delta \lambda_{BWM}$ from equations (\ref{shift_SMW}) and (\ref{shift_BMW}):
\begin{eqnarray}
\label{shift_Lambda_SMW}
\delta \lambda_{SWM}=\frac{4\pi g}{\varphi} \left (\frac{\Delta \lambda }{\lambda_0} \right)^{2}
=\frac{16\pi S\Omega}{\varphi c}  \left (\frac{\Delta \lambda }{\lambda_0} \right)^{2},
\end{eqnarray}
\begin{eqnarray}
\label{shift_Lambda_BMW}
\delta \lambda_{BWM}=\frac{4\pi g}{\varphi}
=\frac{16\pi S\Omega}{\varphi c}  .
\end{eqnarray}

Note that $\lambda_0$ is usually larger than its uncertainty $\Delta \lambda$. Specifically, the scheme of BWM gives a further amplified mean value shift of the photon wavelength $\lambda$, resulting in sensitivity outperforming SWM by the amplification factor $(\lambda_0/\Delta \lambda )^{2}$. However, the pre-coupling progress will generate an additional reduction of photons in the post-selection. Therefore, we calculate the probability $|{\bkt{\Psi_{f}}{\Psi_{i}}}|^{2}$ of successful pot-selection in both the scheme of SWM and the scheme of BWM, which can be obtained from equations (\ref{final_probe_SWM}) and (\ref{final_probe_BWM}):
\begin{eqnarray}
\label{postselection_SMW}
|{\bkt{\Psi_{f}}{\Psi_{i}}}|^{2}_{SWM}={\rm sin^{2}} (gp+\varphi),
\end{eqnarray}
\begin{eqnarray}
\label{postselection_BMW}
|{\bkt{\Psi_{f}}{\Psi_{i}}}|^{2}_{BWM}={\rm sin^{2}} (gp) .
\end{eqnarray}

In order to effectively compare the scheme of SWM and the scheme of BWM, we obtain the numerical results of the two schemes with determinate parameters. Here, the wavelength of the laser is centred at $\lambda_{0}$=833 nm with 20 nm full width at half maximum. The area S= 1000 $m^{2}$ is enclosed by the light path in the Sagnac’s interferometer, which can be realized by a fiber loop. The angular between the optical axes of QWP and P2 is set as $\varphi$=0.0001 rad.

The numerical results of the simulations of SWM and BWM are shown in Figure \ref{result_numberical}. The shifts of the probe of SWM and BWM detected at $\Omega=1.0 \times 10^{-9}$ rad/sec and $\Omega=1.9 \times 10^{-8}$ rad/sec are shown in Figure \ref{result_numberical}(A) and Figure \ref{result_numberical}(B). The change of central wavelength $\delta \lambda_{SWM}$ in the scheme of SWM is more obvious than $\delta \lambda_{BWM}$ in the scheme of BWM. And a more specific comparison is shown in Figure \ref{result_numberical}(C), through SWM, the sensitivity is improved by $(\lambda_0/\Delta \lambda )^{2}\approx40$ compared to that of SWM, which keeps the same at different rotation velocity $\Omega$. Note that the price of the sensitivity enhanced in BWM is reducing the probability $|{\bkt{\Psi_{f}}{\Psi_{i}}}|^{2}$ of successful pot-selection, we find that the effect of reduction gets smaller when the coupling strength(which is proportional to rotation velocity $\Omega$) increase. In addition, the lower probability $|{\bkt{\Psi_{f}}{\Psi_{i}}}|^{2}$ of successful pot-selection may not lead to negative effects, an additional reduction of photons in the post-selection make the detector response difficult to be sutured\cite{2021Yin}.

\section{Conclusions}
We proposed a new scheme for enhancing the sensitivity of amplification of rotation velocity in Sagnac's interferometer with biased weak measurement. By introducing a pre-coupling in the standard weak measurement, our numerical results show that the sensitivity of BWM is improved by $(\lambda_0/\Delta \lambda )^{2}$ compared to that of SWM. The proposed scheme can be applied in various optical weak measurements to enhance the sensitivity and the relevant experiments are underway.

\section*{Acknowledgements}
This study is financially supported by the National Key Research and Development Program of China (Grant No. 2018YFC1503705).


\section*{References}
\begin{thebibliography}{9}
\bibitem{AAV} Y.  Aharonov,  D.  Z.  Albert,   and  L.  Vaidman,  Phys.Rev. Lett. \textbf{60}, 1351 (1988).


\bibitem{PhysRevLett.112.200401} O.S.Maga$\widetilde{n}$a Loaiza, M. Mirhosseini, B. Rodenburg,  and R. W. Boyd, Phys. Rev. Lett. \textbf{112}, 200401 (2014).

\bibitem{PhysRevA.82.063822} D. J. Starling, P. B. Dixon, A. N. Jordan,  and J. C.Howell, Phys. Rev. A \textbf{82}, 063822 (2010).

\bibitem{2013Weak} G. I. Viza, J. Mart$\acute{i}$nez-Rinc$\acute{o}$n, G. A. Howland, H. Frostig, I. Shomroni, B. Dayan, and J. C. Howell, 	Opt. Lett. \textbf{38}, 2949 (2013). 

\bibitem{Santana:19} O. J. S. Santana and L. E. E. de Araujo, J. Opt. Soc. Am. B \textbf{36}, 533 (2019).

\bibitem{Huang2021}  J.H. Huang,  X.Y. Duan,   and X.Y. Hu, Eur. Phys. J. D \textbf{75}, 114 (2021).

\bibitem{Huang2021-2}  J.H. Huang,  X.Y. Duan,  G.J Wang and X.Y. Hu, Eur. Phys. J. D \textbf{75}, 256 (2021).

\bibitem{2021Weak}  S. Li, Z. Chen, L. Xie, Q. Liao,   and X. Lin, 	Opt. Express \textbf{29} ,8777 (2021).

\bibitem{2021Yin} P. Yin, W.-H. Zhang, L. Xu, Z.-G. Liu, W.-F. Zhuang,L. Chen, M. Gong, Y. Ma, X.-X. Peng, G.-C. Li, J.-S.Xu, Z.-Q. Zhou, L. Zhang, G. Chen, C.-F. Li, and G.-C. Guo, Light: Science $\&$ Applications \textbf{10}, 103 (2021).

\bibitem{BWM2016}Z.-H. Zhang, G. Chen, X.-Y. Xu, J.-S. Tang, W.-H. Zhang, Y.-J. Han, C.-F. Li, and G.-C. Guo, Phys. Rev. A \textbf{94} , 053843 (2016).

\bibitem{Post1967} E. Post, Geophys. J. Int.475 (1967)

\bibitem{Heiner2010} I. Heiner, C. Alain, W. Joachim, F. Asher, S. Ulrich, V.
Alex, P. D. Nguyen, Geophys. J. Int. \textbf{1}, 182 (2010).

\bibitem{Qiu2017} X. Qiu, L. Xie, X. Liu, L. Luo, Z. Li, Z. Zhang, and J. Du, Appl.
Phys. Lett.\textbf{110}, 071105 (2017).

\bibitem{Zhu2021} J. Zhu, Z. Li, Y. Liu, Y. Ye, Q. Ti, Z. Zhang, and F. Gao,Phys. Rev. A \textbf{103}, 032212 (2021)

\end{thebibliography}
\end{document}